\documentclass[prl,twocolumn]{revtex4}
\usepackage{graphicx}

\begin{document}

\title{A large magnetic storage ring for Bose-Einstein condensates}

\author{A.~S.~Arnold}
\author{C.~S.~Garvie}
\author{E.~Riis}
 \affiliation{Dept.~of Physics, University of Strathclyde, Glasgow G4 0NG, UK\\
 www.photonics.phys.strath.ac.uk}

\date{\today}

\begin{abstract}
Cold atomic clouds and Bose-Einstein condensates have been stored in a $10\,$cm diameter vertically-oriented magnetic
ring. An azimuthal magnetic field enables low-loss propagation of atomic clouds over a total distance of $2\,$m, with a
heating rate of less than $50\,$nK/s. The vertical geometry was used to split an atomic cloud into two counter-rotating
clouds which were recombined after one revolution. The system will be ideal for studying condensate collisions and
ultimately Sagnac interferometry.
\end{abstract}

\maketitle

%\pacs{03.75.Fi, 03.75.Be, 32.80.-t, 05.60.Gg}

The field of atom optics \cite{erlchaifan} has seen a plethora of
significant advances since the advent of laser cooling
\cite{lascool} and gaseous Bose-Einstein condensation (BEC)
\cite{And1}. Analogs of mirrors, lenses, and beamsplitters for
manipulating ultra-cold atoms now exist and are continually being
improved.

A relatively recent addition to the atom-optical toolbox is the cold atom storage ring,
where atoms are guided around a closed path. This is an interesting configuration for
performing, for instance, ultra-sensitive Sagnac \cite{Sag} atom interferometry. An
electrostatic storage ring was first reported for molecules \cite{Mei} in 2001. However,
for many interferometry experiments atoms are more suitable candidates, as they can be
easily laser cooled and/or prepared in the same quantum state.

\begin{figure}[ht]
\begin{center}\mbox{\includegraphics[clip,width=\columnwidth]{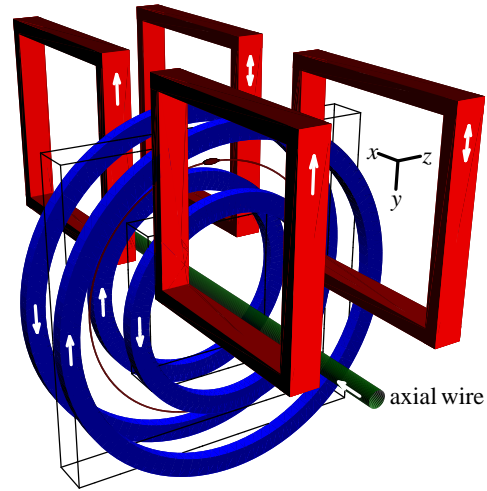}}\end{center}
\vspace{-23mm}\caption{(Color online) Four circular coils with average diameter $10\,$cm
make up the storage ring (thin circle) to which the axial wire adds an adjustable
azimuthal magnetic field. The square coils localise the atoms (small cigar) at the top of
the ring in a magneto-optical (magnetic) trap if the left and right square coil pairs
have the same (opposite) currents.}
 \label{IPcoils}
\end{figure}

The first atomic storage ring was formed with the magnetic
quadrupole field created by two concentric current carrying loops
\cite{sau}. Our storage ring \cite{us} utilises the more symmetric
quadrupole field of a four-loop geometry (Fig.~\ref{IPcoils}). A
third group has recently developed a storage `stadium' \cite{prent}
based on a magnetic waveguide. Our ring contains more than
$5\times10^8$ atoms, has a lifetime of $50\,$s and an area of
$72\,\rm{cm}^2,$ each corresponding to more than an order of
magnitude improvement with respect to Refs. \cite{sau,prent}. The
Sagnac effect \cite{Sag} is linearly proportional to the area of an
interferometer, and our ring's area compares favourably with that of
a state-of-the-art thermal beam atom interferometer gyro
$(A=0.22\,\rm{cm}^2$ \cite{thermgyro}). The atoms can complete
multiple revolutions in the ring, further increasing our effective
area and thus the rotation sensitivity. In addition, our storage
ring has the unique feature that we are able to form a BEC in a
section of the ring~\cite{us}, and observe its propagation around
the ring.

We begin by considering the storage ring theoretically, before discussing our experimental setup.
Comparisons will then be drawn between our experiment and theory.

All magnetic atom-optical elements make use of the Stern-Gerlach potential $U=\mu_{B}
g_{F} m_{F} B$ experienced by an atom moving adiabatically in a magnetic field of
magnitude $B$, where $\mu_{B}$ is the Bohr magneton, $m_{F}$ the atom's hyperfine
magnetic quantum number, and $g_{F}$ is the Land\'{e} g-factor. To make a storage ring
one must use atoms in weak-field-seeking magnetic states $(g_{F} m_{F}>0),$ which are
attracted to minima of the magnetic field magnitude. Using the Biot-Savart law, one can
express the cylindrically symmetric magnetic field from a single coil of radius $R,$
$\textbf{b}_R(r,z),$ in terms of elliptic integrals \cite{good}. Our storage ring
comprises four concentric circular coils (Fig.~\ref{IPcoils}), with a toroidal quadrupole
total magnetic field:
\begin{eqnarray}
\textbf{B}(r,z)&=&\textbf{b}_{R-\delta_R}(r,z-\delta_z)-\textbf{b}_{R-\delta_R}(r,z+\delta_z)\nonumber\\
               &-&\textbf{b}_{R+\delta_R}(r,z-\delta_z)+\textbf{b}_{R+\delta_R}(r,z+\delta_z).
\label{ringfield}
\end{eqnarray}
In our experiment $R=5.0\,\rm{cm},$ $\delta_R=1.25\,\rm{cm},$ and
$\delta_z=1.35\,\rm{cm},$ leading to a ring of zero magnetic field at a radius
$R_0=4.8\,\rm{cm}$ slightly smaller than the mean coil radius $R.$ The axial wire
(Fig.~\ref{IPcoils}) adds a $1/r$ azimuthal magnetic field to Eq.~\ref{ringfield}, which
has little effect on the ring radius, but yields a storage ring with non-zero magnetic
field.

The theory of the storage ring for cold atoms is relatively simple, as the
spatio-temporal evolution of each atom can be accurately described using the equations of
motion arising from the Stern-Gerlach potential. After choosing suitable Gaussian initial
position and velocity distributions it is possible to perform a Monte Carlo simulation to
build up a spatio-temporal atomic density map.

\begin{figure}[ht]
\begin{center}\mbox{\includegraphics[clip,width=1\columnwidth]{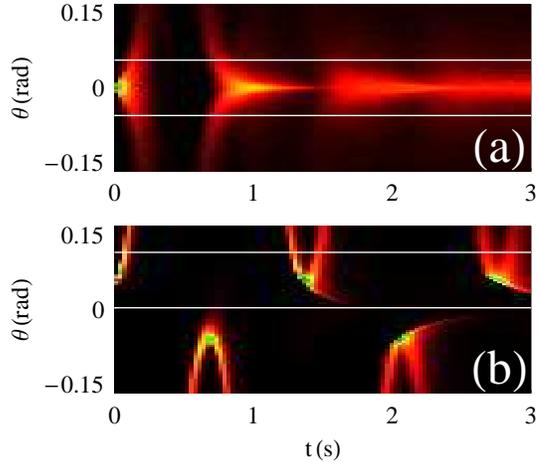}}\end{center}
\vspace{-7mm}\caption{(Color online) A Monte-Carlo simulation of atomic dynamics in the
vertically-oriented storage ring for a $2\,\mu$K initial cloud of atoms. Images (a) and
(b) represent lossless atom dynamics for a cloud released at $\overline{\theta_0}=0$ and
$\overline{\theta_0}=60\,$mrad respectively. The white lines mark the extent of our
$110\,$mrad experimental viewing region.}
 \label{MC}
\end{figure}

By comparison with a 3D model, we have found that a 1D model is
sufficient, i.e. the rigid pendulum equation: \begin{equation}
\theta''(t)=(g/R_0)\sin\theta, \label{rigpen}
\end{equation}
where $\theta$ and $\omega=\frac{d\theta}{dt}$ are the angular
position and velocity of an atom, $g$ is the acceleration due to
gravity and $R_0$ is the radius of the storage ring. Note that
Eq.~\ref{rigpen} can be integrated analytically (using
$\frac{d^2\theta}{dt^2}=\omega \frac{d\omega}{d\theta})$ given the
initial angle $(\theta_0)$ and angular velocity $(\omega_0)$ of an
atom, to determine the time-averaged relative probability of finding
an atom at a given angle:
\begin{equation}
P(\theta)\propto
1/\omega=1/\sqrt{{\omega_0}^2+2g(\cos\theta_0-\cos\theta)/R_0}.
\label{rigpena}
\end{equation}
In our vertically-oriented storage ring there is therefore a high time-averaged
probability of finding atoms in the region where they travel the slowest -- the top of
the ring. The integral of Eq.~\ref{rigpena}, leads to an expression for $t(\theta)$ in
terms of an elliptic integral, which can be inverted to find $\theta(t)$. There are two
kinds of trajectories in the ring: if
$\cos(\theta_{max})=|\frac{R_0}{2g}{\omega_0}^2+\cos\theta_0|>1$ an atom will always
rotate in the same direction, but if $\cos(\theta_{max})\leq 1$ (cf.~Fig.~\ref{MC}(b))
the atom will reverse its direction around the ring at the turning points
$\theta=\pm\theta_{max}$.

The theoretical time-dependent angular distribution of atoms can be
seen in Fig.~\ref{MC}. An atomic cloud centered in a parabolic
potential can expand or shrink, but will maintain the same cloud
shape. A circular atomic trajectory has a potential which is only
parabolic to second order, and it is the higher order effects which
cause an atomic cloud released from the top of the ring ($\theta$=0)
to break into two halves (Fig.~\ref{MC}(a)), leaving a near-zero
probability at the top of the ring after around $300\,$ms. These two
halves return to the top of the ring after a further $300\,$ms and
have a non-zero probability of remaining at the top of the ring for
all subsequent times. The time for the two halves of the atomic
cloud to `recombine' has only a weak dependence on the initial
atomic temperature. Gravity is the dominant effect for cold atoms,
leading to a `hot' velocity of $1.4\,\rm{m}\,\rm{s}^{-1}$ at the
bottom of the ring -- a good location for future studies of
high-energy ($20\,$mK) collisions between BECs \cite{WilWal}.

Although the spread of initial thermal velocities quickly ensures that atoms reach the
top of the ring at different times, in Fig.~\ref{MC}(a) we still see slowly decaying
`echos' of the originally localised spatial atomic cloud in the form of time-varying
bimodal perturbations of the $\theta$ distribution. The angular width of these echos is
approximately proportional to the initial cloud width, so the echos and their bimodal
nature become more pronounced at higher temperatures.

In many of our ring experiments the atoms/BECs were not launched from $\overline{\theta_0}=0$ (Fig.~\ref{MC}(a)).
Although the total number of atoms observed is similar to the $\overline{\theta_0}\neq 0$ case, the size of the returning
$\overline{\theta_0}\neq 0$ cloud is much smaller and clearer due to focusing at the turning points of its motion
(Fig.~\ref{MC}(b)).

We now turn to the experimental setup, for which many of the details
are in Refs.~\cite{us,AARIIS}. Our double magneto-optical trap (MOT)
\cite{dmot} collects $10^9$ $^{87}$Rb atoms in both our high and low
pressure MOT chambers. The storage ring can be loaded directly from
the low pressure MOT or Ioffe-Pritchard (IP \cite{IP}) trap without
the need for any additional magnetic guiding or transfer, i.e. the
MOT and IP trap are at the top of the ring. Releasing atoms at the
top of a vertical ring ensures insensitivity to magnetic ring
corrugations and a fast rotation frequency.

The advantage of our hybrid magnetic trap (Fig.~\ref{IPcoils}) is
that, using the same coils, it can be (i) a MOT, (ii) an IP magnetic
trap (with trap frequencies $\nu_r=230\,\rm{Hz},$
$\nu_z=10\,\rm{Hz})$ or iii) a toroidal magnetic storage ring with
radial gradient $230\,$G/cm. In modes (i) and (ii) the magnetic
field has an adjustable aspect ratio in the azimuthal direction. The
four 2~turn$\times500\,$A circular coils in Fig.~\ref{IPcoils}
create a toroidal magnetic quadrupole field, confining the atoms to
a ring. The four 3~turn$\times500\,$A square `pinch' coils are wired
in pairs, and confine the atoms to a localised section of the ring
for Ioffe-Pritchard trapping of the atoms.  An azimuthal field can
be added to the storage ring via an axial wire. The large coil sizes
ensure good optical access, and low magnetic field noise. The
absorption imaging (gravity) direction is the $x$ $(y)$ axis.

After loading the low pressure MOT, the atomic cloud is optically
pumped, loaded into the IP trap, magnetically compressed, and then
evaporatively cooled to an adjustable temperature. The atoms can
then be smoothly loaded into the storage ring in $20\,$ms. We are
able to form Bose-Einstein condensates containing $N_0=2\times 10^5$
$|F=2,m_F=2\rangle$ atoms at the top of the storage ring \cite{us}
at a typical final RF evaporation frequency of $750\,\rm{kHz}$.

\begin{figure}[hb]
\begin{center}
\vspace{-4mm}\mbox{\includegraphics[clip,width=1\columnwidth]{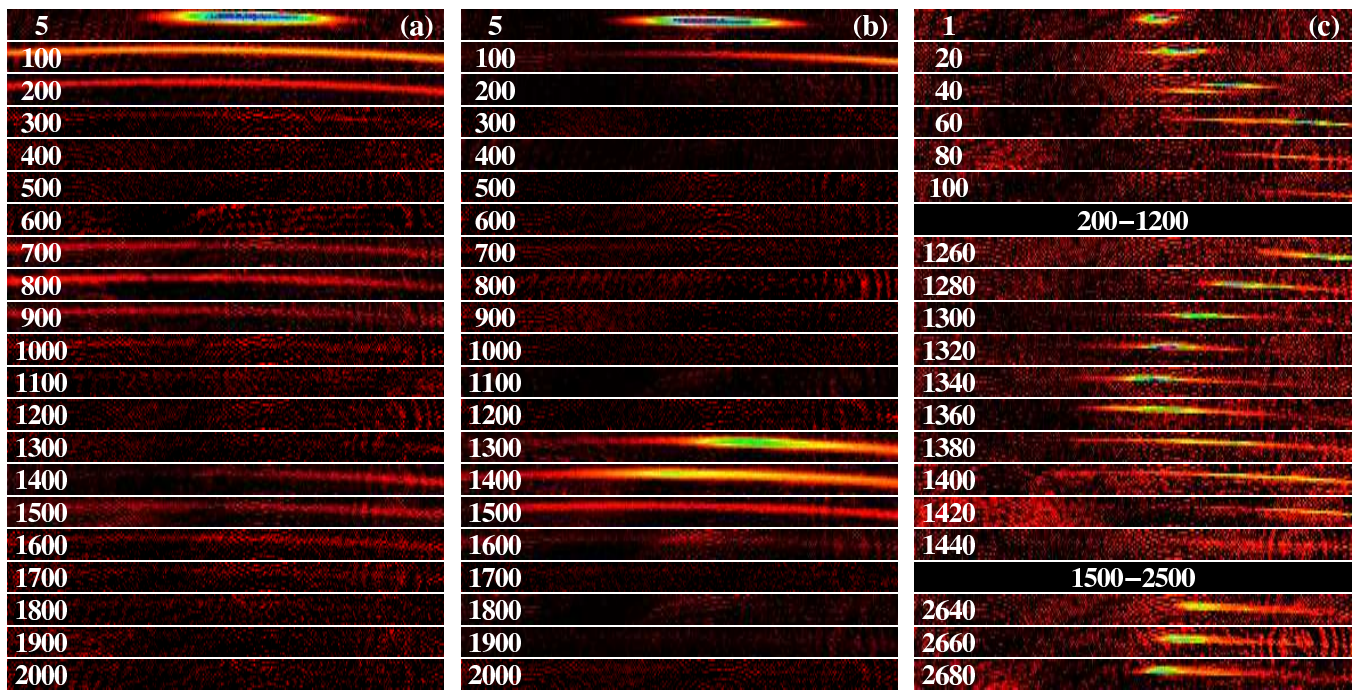}}\end{center}
 \vspace{-4mm}\caption{(Color online) Experimental storage ring dynamics for 2$\,\mu$K atoms with (a)
$B_\theta=0\,$G and (b) $B_\theta=10\,$G. In (c) multiple
revolutions of a BEC in a ring with $B_\theta=10\,$G are seen.
Numbers denote time in ms, absorption images are $0.4\times
5.4\,\rm{mm}^2$. \label{6Ring}}
\end{figure}

In order to compare our experimental ring data (Fig.~\ref{6Ring})
with our theoretical model (Fig.~\ref{MC}), we have found it
convenient to model the total number of atoms in the `viewing
window' of our absorption imaging system as a function of time. Our
CCD camera has an area of $4.8\times6.4\,\mbox{mm}^2,$ with a
magnification of $1.20(1).$ We have released cold atomic clouds with
a variety of final evaporation temperatures into a storage ring with
no azimuthal field. In accordance with theory, the atoms disappear
after $300-400\,$ms, reappear shortly afterwards, and are present at
all subsequent times. However, there is a marked variation in the
storage ring dynamics with atomic cloud release temperature. This
can be explained in terms of non-adiabatic Majorana spin-flip
transitions \cite{TOP}. The storage ring has a ring of zero magnetic
field, and colder atoms pass closer to the magnetic field zero, and
are selectively removed from the storage ring.

If we apply a constant magnetic field across the entire storage ring, perpendicular to
the ring axis, it is possible to have only two places in the ring with zero magnetic
field, however there will be a strong angular variation in the trapping potential. A
novel feature of our storage ring is that the `hole' in our quartz vacuum chamber
(Fig.~\ref{IPcoils}) allows us to use a wire along the axis of the storage ring. This
means that we can generate an azimuthal magnetic bias field $B_{\theta}=0-10\,$G around
the ring which transforms the radial magnetic potential from a cone to a hyperbola,
removing the ring of zero magnetic field and drastically reducing atomic loss.

\begin{figure}[ht]
\begin{center}\mbox{\includegraphics[clip,width=1\columnwidth]{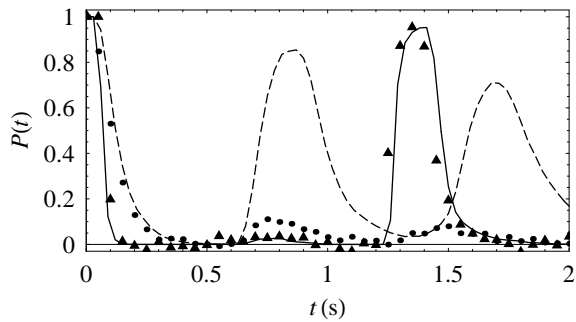}}\end{center}
\vspace{-5mm}\caption{The relative atomic population, $P(t),$ viewed
by the CCD camera as a function of time for 2$\,\mu$K atoms.
Experimental data with $B_\theta=0\,$G and $B_\theta=10\,$G are
shown with circles and triangles respectively. The relative Monte
Carlo atomic population in the viewing region of Fig.~\ref{MC}(a)
(dashed curve) and Fig.~\ref{MC}(b) (solid curve) are shown for
comparison. Good quantitative agreement can be seen between the
$B_\theta=10\,$G data and the Monte Carlo simulation of an atomic
cloud released from $\theta_0=60\,$mrad. The $B_\theta=0\,$G data
only fits the expected dashed theoretical curve initially due to
large Majorana losses.} \label{Ntgraph}
\end{figure}

This difference can be seen in the experiments of
Fig.~\ref{6Ring}(a) and (b), where $2\,\mu$K atomic clouds are
released into a ring without and with an azimuthal field,
respectively. A quantitative contrast of the number of visible atoms
in each case is shown in Fig.~\ref{Ntgraph}, as well as a comparison
to the Monte Carlo theory of Fig.~\ref{MC}. The difference the
azimuthal field makes is even greater with condensates: without it
BECs vanish before completing one revolution, but with an azimuthal
field multiple revolutions of a condensate are possible with low
loss (Fig.~\ref{6Ring}(c)).

\vspace{-4mm} \onecolumngrid\widetext
\twocolumngrid

Note that when we performed experiments with an azimuthal field, the atoms/BECs were not
launched from $\overline{\theta_0}=0.$ If atoms are launched from $\overline{\theta_0}=0$
(Fig.~\ref{MC}(a)), then although the total number of atoms observed (Fig.~\ref{Ntgraph})
is similar to the $\overline{\theta_0}\neq 0$ case, the size of the returning
$\overline{\theta_0}\neq 0$ cloud is much smaller and clearer due to focusing in the ring
(Figs.~\ref{MC}(b),~\ref{6Ring}(b,c)).

We do not expect to see phase-fluctuations \cite{phasefluc} in our condensate before or
after propagation in the ring. These effects have been studied in highly elongated
condensates in which the BEC coherence length is less that the length of the condensate.
We form the condensate in only a moderately elongated trap, and the azimuthal expansion
process is rapid enough that the (density-dependent) phase fluctuations do not have time
to develop.

\begin{figure}[ht]
\begin{center}\mbox{\includegraphics[clip,width=1\columnwidth]{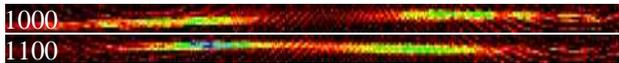}}\end{center}
\vspace{-5mm}\caption{A BEC at the top of the ring was split into two counter-rotating
clouds by the faster-than-harmonic circular potential. These split atomic clouds can be
seen recombining here after one revolution in the ring. Times are in ms, each image is
$150\times3400\,\mu\rm{m}^2$ and is taken after a $3\,$ms ballistic expansion.}
\label{split}
\end{figure}

Sagnac interferometry will be performed by locating the condensate at the exact top of
the ring, and incoherently splitting the sample by simply releasing it
(Fig.~\ref{split}). We are currently looking for interference fringes after a single
revolution. Coherent splitting of the BEC will be achieved using Bragg scattering
\cite{bragg} to send BEC wavepackets in both directions around the ring. Note that
ideally the phase sensitivity of a BEC interferometer scales like
$\delta\phi_{BEC}\propto N^{-1}$ where $N$ is the number of atoms (cf.
$\delta\phi_{T}\propto N^{-1/2}$ for thermal atoms), however the increased sensitivity is
only possible if the BECs are prepared in number states \cite{yama}.

In conclusion, we have demonstrated a storage ring for cold
$^{87}$Rb atomic clouds and BECs with an area of $7200\,\rm{mm}^2$.
An azimuthal bias field around the ring enables low-loss BEC
propagation with heating of less than $50\,$nK/s. Our goal is a
highly sensitive Sagnac atom interferometer, in which we are aided
by our unprecedented ring area. Rotation sensitivity for a single
revolution of $\Delta\Omega=\hbar/(8 m \pi
{R_0}^2\sqrt{N})=3\,10^{-11}\rm{rad/s}$ is feasible.

If one cannot create an azimuthal magnetic field, it is still
possible to prevent spin-flip losses by using an adjustable-radius
time-orbiting ring trap (TORT) \cite{TORT}. We recently learnt that
a team in Berkeley has created the first TORT for BECs
\cite{stamper}, with ring area $5\,\rm{mm}^2$. Large BECs were also
formed in an optically-plugged ring trap with ring area $\ll
1\,\rm{mm}^2$ \cite{raman}.

We are grateful for helpful discussions with K.~Burnett and E.~Hinds. This work was
supported by the UK EPSRC and the University of Strathclyde.

\end{document}